\documentclass[twocolumn,english,english,reprint,prl,aps,superscriptaddress]{revtex4-1}
\usepackage[T1]{fontenc}
\usepackage[latin9]{inputenc}
\usepackage{geometry}
\usepackage{array}
\usepackage{float}
\usepackage{multirow}
\usepackage{amsmath}
\usepackage{amssymb}
\usepackage{graphicx}

\makeatletter

\providecommand{\tabularnewline}{\\}

\@ifundefined{textcolor}{}
{%
 \definecolor{BLACK}{gray}{0}
 \definecolor{WHITE}{gray}{1}
 \definecolor{RED}{rgb}{1,0,0}
 \definecolor{GREEN}{rgb}{0,1,0}
 \definecolor{BLUE}{rgb}{0,0,1}
 \definecolor{CYAN}{cmyk}{1,0,0,0}
 \definecolor{MAGENTA}{cmyk}{0,1,0,0}
 \definecolor{YELLOW}{cmyk}{0,0,1,0}
}

\makeatother

\usepackage{babel}
\begin{document}

\title{Breaking the theoretical scaling limit for predicting quasi-particle
energies: The stochastic GW approach}

\author{Daniel Neuhauser}

\affiliation{Department of Chemistry, University of California at Los Angeles,
CA-90095 USA}

\author{Yi Gao}

\affiliation{Department of Chemistry, University of California at Los Angeles,
CA-90095 USA}

\author{Christopher Arntsen}

\affiliation{Department of Chemistry, University of California at Los Angeles,
CA-90095 USA}

\author{Cyrus Karshenas}

\affiliation{Department of Chemistry, University of California at Los Angeles,
CA-90095 USA}

\author{Eran Rabani}

\affiliation{School of Chemistry, The Sackler Faculty of Exact Sciences, Tel Aviv
University, Tel Aviv 69978, Israel}

\author{Roi Baer}

\affiliation{Fritz Haber Center for Molecular Dynamics, Institute of Chemistry,
The Hebrew University of Jerusalem, Jerusalem 91904, Israel}
\begin{abstract}
We develop a formalism to calculate the quasi-particle energy within
the GW many-body perturbation correction to the density functional
theory (DFT). The occupied and virtual orbitals of the Kohn-Sham (KS)
Hamiltonian are replaced by stochastic orbitals used to evaluate the
Green function, the polarization potential, and thereby the GW self-energy.
The stochastic GW (sGW) relies on novel theoretical concepts such
as stochastic time-dependent Hartree propagation, stochastic matrix
compression and spatial/temporal stochastic decoupling techniques.
Beyond the theoretical interest, the formalism enables \textit{linear
scaling} GW calculations breaking the theoretical scaling limit for
GW as well as circumventing the need for energy cutoff approximations.
We illustrate the method for silicon nanocrystals of varying sizes
with $N_{e}>3000$ electrons.
\end{abstract}
\maketitle
The GW approximation~\cite{Hedin1965,Friedrich2006} to many-body
perturbation theory (MBPT)~\cite{Fetter1971} offers a reliable and
accessible theory for quasi-particles (QPs) and their energies \cite{Hybertsen1985,Hybertsen1986,Aulbur2000,Stan2006,Friedrich2006,Shishkin2007,Huang2008,Trevisanutto2008,Rostgaard2010,Liao2011,Berger2012,Setten2012,Marom2012,Isseroff2012,Refaely-Abramson2012,Caruso2013},
enabling estimation of electronic excitations~\cite{Grossman2001,Onida2002,Tiago2006,Refaely-Abramson2011,Caruso2012,Blase2011a,Faleev2004}
quantum conductance~\cite{Quek2007,Quek2007a,Thygesen2008,Myohanen2009,DellaSalla2010}
and level alignment in hybrid systems~\cite{Neaton2006,Tamblyn2011}.
Practical use of GW for large systems is severely limited because
of the steep CPU and memory requirements as system size increases.
The most computationally intensive element in the GW method, the calculation
of the polarization potential (screen Coulomb interaction), involves
an algorithmic complexity that scales as the fourth power of the system
size~\cite{Nguyen2012,Deslippe2012}. Various approaches have been
developed to reduce the computational bottlenecks of the GW approach~\cite{Deslippe2012,Nguyen2012,Pham2013,Shishkin2007,Foerster2011,Gonze2009,Caruso2013,Caruso2012}.
Despite these advances, GW calculations are still quite expensive
for many of the intended applications in the fields of materials science,
surface science and nanoscience.

In this letter we develop a stochastic, orbital-less, formalism for
the GW theory, unique in that it does not reference occupied or virtual
orbitals and orbital energies of the KS Hamiltonian. While the approach
is inspired by recent developments in electronic structure theory
using stochastic orbitals~\cite{Baer2012a,Neuhauser2013,Neuhauser2013a,Baer2013,Neuhauser2014}
it introduces three powerful and basic notions: Stochastic decoupling,
stochastic matrix compression and stochastic time-dependent Hartree
(sTDH) propagation. The result is a stochastic formulation of GW,
where the QP energies become random variables sampled from a distribution
with a mean equal to the exact GW energies and a statistical error
proportional to the inverse number of stochastic orbitals (iterations,
$I_{sGW}$). 

We illustrate the sGW formalism for silicon nanocrystals (NCs) with
varying sizes and band gaps~\cite{Wang1994d,Ogut1997} and demonstrate
that the CPU time and memory required by sGW scales \textit{nearly
linearly} with system size, thereby providing means to study QPs excitations
in large systems of experimental and technological interest.

In the reformulation of the GW approach, we treat the QP energy ($\varepsilon_{QP}=\hbar\omega_{QP}$)
as a perturbative correction to the KS energy~\cite{Hybertsen1986,Friedrich2006}:
\begin{equation}
\varepsilon_{QP}\left(\varepsilon\right)=\varepsilon+\tilde{\Sigma}^{P}\left(\omega_{QP};\varepsilon\right)+\Sigma^{X}\left(\varepsilon\right)-\Sigma^{XC}\left(\varepsilon\right).\label{eq:eps-QP}
\end{equation}
We view the KS energy $\varepsilon$ as a variable (rather than an
eigenvalue) and the actual value we use is determined from the density
of states of the KS Hamiltonian available from the sDFT calculation~\cite{Baer2013}.
For each value of $\varepsilon$ one needs to evaluate the self-energy
in Eq.~\eqref{eq:eps-QP} given by the sum of the self-energy terms:
\begin{equation}
\begin{array}{rcl}
\Sigma^{P}\left(t;\varepsilon\right) & = & \frac{1}{Q\text{\ensuremath{\left(\varepsilon\right)}}}tr\left[f_{\sigma}\left(\hat{h}_{KS}-\varepsilon\right)^{2}\hat{\Sigma}^{P}\left(t;\varepsilon\right)\right],\\
\\
\Sigma^{X}\left(\varepsilon\right) & = & \frac{1}{Q\text{\ensuremath{\left(\varepsilon\right)}}}tr\left[f_{\sigma}\left(\hat{h}_{KS}-\varepsilon\right)^{2}\hat{\Sigma}^{X}\right],\\
\\
\Sigma^{XC}\left(\varepsilon\right) & = & \frac{1}{Q\text{\ensuremath{\left(\varepsilon\right)}}}tr\left[f_{\sigma}\left(\hat{h}_{KS}-\varepsilon\right)^{2}v_{XC}\right].
\end{array}\label{eq:SE<trace>}
\end{equation}
The frequency domain polarization self-energy $\tilde{\Sigma}^{P}\left(\omega,\varepsilon\right)$
is given in terms of the Fourier transform of the time domain counterpart
$\Sigma^{P}\left(t,\varepsilon\right)$. $\Sigma^{X}\left(\varepsilon\right)$
and $\Sigma^{XC}\left(\varepsilon\right)$ are the exchange and exchange-correlation
self-energies, respectively, and $Q\text{\ensuremath{\left(\varepsilon\right)}}=tr\left[f_{\sigma}\left(\hat{h}_{KS}-\varepsilon\right)^{2}\right]$
is a normalization factor. In the above, $v_{XC}\left(\mathbf{r}\right)$
is the exchange-correlation potential of the KS-DFT Hamiltonian $\hat{h}_{KS}$
and $f_{\sigma}(\varepsilon)=e^{-\varepsilon^{2}/2\sigma^{2}}$ is
an energy filter function of width $\sigma$~\cite{Neuhauser1990}.
$\Sigma^{X}\left(\varepsilon\right)$, $\Sigma^{XC}\left(\varepsilon\right)$,
and $Q\text{\ensuremath{\left(\varepsilon\right)}}$ can be calculated
using a linear-scaling stochastic approach, as detailed in the supplementary
information.

In the GW approximation, the most demanding calculation involves the
polarization self-energy, formally given by~\cite{Friedrich2006}:
\begin{equation}
\begin{array}{ll}
\Sigma^{P}\left(\mathbf{r}_{1},\mathbf{r}_{2},t;\varepsilon\right) & =\left\langle \mathbf{r}_{1}\left|\hat{\Sigma}^{P}\left(t;\varepsilon\right)\right|\mathbf{r}_{2}\right\rangle =\\
 & i\hbar G_{0}\left(\mathbf{r}_{1},\mathbf{r}_{2},t\right)W^{P}\left(\mathbf{r}_{1},\mathbf{r}_{2},t;\varepsilon\right),
\end{array}\label{eq:GW Approx}
\end{equation}
where 

\begin{equation}
\begin{array}{ll}
i\hbar G_{0}\left(\mathbf{r}_{1},\mathbf{r}_{2},t\right) & \equiv\left\langle \mathbf{r}_{1}\left|e^{-i\hat{h}_{KS}t/\hbar}\hat{P}_{\mu}\left(t\right)\right|\mathbf{r}_{2}\right\rangle ,\end{array}\label{eq:G_0}
\end{equation}
is the Green function and 

\begin{equation}
\begin{array}{ll}
W^{P}\left(\mathbf{r}_{1},\mathbf{r}_{2},t;\varepsilon\right) & \equiv\left\langle \mathbf{r}_{1}\left|u_{C}\otimes\chi\left(t;\varepsilon\right)\otimes u_{C}\right|\mathbf{r}_{2}\right\rangle \end{array}\label{eq:WP}
\end{equation}
is the polarization potential. In the above equations, $\hat{P}_{\mu}\left(t\right)\equiv\left(\theta\left(t\right)-\theta_{\beta}\left(\mu-\hat{h}_{KS}\right)\right)$,
$\theta\left(t\right)$ and $\theta_{\beta}\left(E\right)=\frac{1}{2}\left(1+\mbox{erf}\left(\beta E\right)\right)$
are the Heaviside and a smoothed-Heaviside functions, respectively,
$\mu$ is the chemical potential, $u_{C}\left(\left|\mathbf{r}_{1}-\mathbf{r}_{2}\right|\right)=e^{2}/4\pi\epsilon_{0}\left|\mathbf{r}_{1}-\mathbf{r}_{2}\right|$
is the bare Coulomb potential, and $\chi\left(\mathbf{r}_{1},\mathbf{r}_{2},t;\varepsilon\right)$
is the \emph{time-ordered }density-density correlation function~\cite{Fetter1971}.
The symbol $'\otimes'$ represents a space convolution. 

Instead of performing the trace operations in Eqs.~(\ref{eq:SE<trace>})-(\ref{eq:WP})
using the full basis of $\hat{h}_{KS}$, which for large system is
prohibitive, we use a relatively small set of \textit{real }stochastic
orbitals $\phi\left(\mathbf{r}\right)$~\cite{Drabold1993,Wang1994b,silver1996chebyshev}
for which $\mathbf{1}=\left<\left|\phi\left\rangle \right\langle \phi\right|\right>_{\phi}$
where $\left<\cdots\right>_{\phi}$ denotes a statistical average
over $\phi$. The choice of $\phi\left(\mathbf{r}\right)$ satisfying
these requirements is not unique. The form used here assigns a value
of $\pm h^{-3/2}$ at each grid point with equal probability, where
$h$ is the grid spacing. This is a crucial step which allows us to
rewrite the self-energy in Eq.~(\ref{eq:SE<trace>}) as:

\begin{equation}
\Sigma^{P}\left(t;\varepsilon\right)=\left<\iint\phi_{\varepsilon}\left(\mathbf{r}_{1}\right)\Sigma^{P}\left(\mathbf{r}_{1},\mathbf{r}_{2},t;\varepsilon\right)\phi\left(\mathbf{r}_{2}\right)d^{3}r_{1}d^{3}r_{2}\right>_{\phi}\label{eq:SE^P}
\end{equation}
where $\left|\phi_{\varepsilon}\right\rangle =f_{\sigma}\left(\hat{h}_{KS}-\varepsilon\right)\left|\phi\right\rangle $
is the corresponding filtered state at energy $\varepsilon$, which
can be obtained by a Chebyshev expansion of the Gaussian function
with $\sigma$ chosen as a small parameter~\cite{Kosloff1988,Kosloff1994}.
We note in passing that it is possible to obtain simultaneously $\Sigma^{P}\left(t;\varepsilon\right)$
for several values of $\varepsilon$ (more details are given in the
supplementary information). 

To obtain $\Sigma^{P}\left(\mathbf{r}_{1},\mathbf{r}_{2},t;\varepsilon\right)$
in Eq.~(\ref{eq:SE^P}) we need to calculate the non-interacting
Green function $i\hbar G_{0}\left(\mathbf{r}_{1},\mathbf{r}_{2},t\right)$
in Eq.~(\ref{eq:G_0}) and the polarization potential $W^{P}\left(\mathbf{r}_{1},\mathbf{r}_{2},t;\varepsilon\right)$
in Eq.~(\ref{eq:WP}). For the former, we introduce an additional
set of real stochastic orbitals, $\zeta\left(\mathbf{r}\right)$,
and describe a:
\begin{equation}
\begin{array}{c}
i\hbar G_{0}\left(\mathbf{r}_{1},\mathbf{r}_{2},t\right)=\left<\zeta_{\mu}\left(\mathbf{r}_{1},t\right)\zeta\left(\mathbf{r}_{2}\right)\right>_{\zeta}\end{array},\label{eq:G_0<stoch>}
\end{equation}

\noindent where $\zeta_{\mu}\left(\mathbf{r},t\right)=\left<\mathbf{r}\left|e^{-i\hat{h}_{KS}t/\hbar}\hat{P}_{\mu}\left(t\right)\right|\zeta\right>$
is a ``propagated-projected'' stochastic orbital which can be obtained
by a Chebyshev expansion of the function $e^{-i\varepsilon t/\hbar}\left(\theta\left(t\right)-\theta_{\beta}\left(\varepsilon-\mu\right)\right)$~\cite{Kosloff1988,Kosloff1994}.
One appealing advantage of the stochastic form of Eq.~(\ref{eq:G_0<stoch>})
is that it provides a compact representation for $G_{0}\left(\mathbf{r}_{1},\mathbf{r}_{2},t\right)$,
equivalent to\emph{ matrix compression} where \textbf{$\mathbf{r}_{1}$}
and $\mathbf{r}_{2}$ are decoupled. This allows a drastic simplification
of the representation of the polarization self-energy obtained by
combining Eqs.~(\ref{eq:SE^P}) and (\ref{eq:G_0<stoch>}):
\begin{equation}
\begin{array}{rcl}
\Sigma^{P}\left(t;\varepsilon\right) & = & \left\langle \left<\phi_{\varepsilon}\zeta_{\mu}\left(t\right)^{*}\left|u_{C}\otimes\chi\left(t\right)\otimes u_{C}\right|\zeta\phi\right>\right\rangle _{\phi\zeta}\end{array}.\label{eq:SE_^P<stoch>}
\end{equation}

\noindent Next, we employ a temporal decoupling scheme achieved by
introducing an additional set of real stochastic orbitals $\psi\left(\mathbf{r}\right)$:

\noindent 
\begin{equation}
\begin{array}{rcl}
\Sigma^{P}\left(t;\varepsilon\right) & = & \left\langle \left<\left.\phi_{\varepsilon}\zeta_{\mu}\left(t\right)^{*}\right|\psi\right>\left<\psi\left|u_{C}\otimes\chi\left(t\right)\otimes u_{C}\right|\zeta\phi\right>\right\rangle _{\phi\zeta\psi}\end{array},\label{eq:SE^P<stoch><decoup>}
\end{equation}

\noindent which\emph{ }allows us to treat the term $\left<\left.\phi_{\varepsilon}\zeta_{\mu}\left(t\right)^{*}\right|\psi\right>$
separately from the term $\left<\psi\left|u_{C}\otimes\chi\left(t\right)\otimes u_{C}\right|\zeta\phi\right>$.
Note that the average $\left\langle \cdots\right\rangle _{\phi\zeta\psi}$
in Eq.~(\ref{eq:SE^P<stoch><decoup>}) is performed over $I_{sGW}$
pairs of $\phi$ and $\zeta$ stochastic orbitals, and for each such
pair we use a different set of $N_{\psi}$ stochastic $\psi's$. The
term $\left<\left.\phi_{\varepsilon}\zeta_{\mu}\left(t\right)^{*}\right|\psi\right>$
is straightforward to obtain while $\left<\psi\left|u_{C}\otimes\chi\left(t\right)\otimes u_{C}\right|\zeta\phi\right>$
is determined from the time-retarded\emph{ }polarization potential,
$\left<\psi\left|u_{C}\otimes\chi^{r}\left(t\right)\otimes u_{C}\right|\zeta\phi\right>$,
calculated from the linear response relation:

\noindent 
\begin{equation}
\left<\psi\left|u_{C}\otimes\chi^{r}\left(t\right)\otimes u_{C}\right|\zeta\phi\right>=\left<\psi\left|u_{C}\right|\delta n\left(t\right)\right>,\label{eq:TDH}
\end{equation}

\noindent where $\delta n\left(\mathbf{r},t\right)$ is the causal
density response to the impulsive perturbation $\delta v\left(\mathbf{r},t\right)=\left<\mathbf{r}\left|u_{C}\right|\zeta\phi\right>\delta(t)$
calculated by the time-dependent Hartree (TDH) approach~\cite{Baroni2001,Baer2004b,Neuhauser2005}.
Alternatively, a full time-dependent density functional theory (TD-DFT)~\cite{Runge1984}
is often found to yield better QP energies than the TDH propagation~\cite{Tiago2006}.
Once the retarded response, $\left<\psi\left|u_{C}\right|\delta n\left(t\right)\right>$,
is calculated and stored for each time $t$, the corresponding time-ordered
response $\left<\psi\left|u_{C}\otimes\chi\left(t\right)\otimes u_{C}\right|\zeta\phi\right>$
is obtained by a standard transformation~%
\footnote{See Fetter and Walecka \cite{Fetter1971} Eq. (13.20).%
}. 

\noindent The TDH (or TD-DFT) propagation is usually performed using
the full set of occupied KS eigenfunctions, but we deliberately avoid
these in our formulation. Instead, we introduce, once again, a \emph{stochastic}
way to perform the TDH or TD-DFT propagation where a new set of $N_{\varphi}$\emph{
}occupied projected stochastic orbitals, $\varphi_{\mu}\left(\mathbf{r},0\right)=\left<\mathbf{r}|\theta\left(\mu-\hat{h}_{KS}\right)|\varphi\right>$
are used (as before $\varphi\left(\mathbf{r}\right)$ are real random
orbitals for which $\mathbf{1}=\left<\left|\varphi\left\rangle \right\langle \varphi\right|\right>_{\varphi}$).
The so called sTDH (or sTD-DFT) propagation is carried out identically
to a TDH propagation, except that one propagates only the $N_{\varphi}$
stochastic orbitals and at each time step (rather than all occupied
orbitals), and the density is calculated as $n\left(\mathbf{r},t\right)=\left\langle \left|\varphi\left(\mathbf{r},t\right)\right|^{2}\right\rangle _{\varphi}$
from which the Hartree potential is updated in the usual way (see
supplementary information for additional details). We verified that
for a given accuracy the number of propagated orbitals $N_{\varphi}$
does not increase (and actually somewhat decreases) with system size~%
\footnote{\noindent For example, in $\mbox{S\ensuremath{i_{35}H_{36}}}$ the
calculated quasi-electron energy is -1.93eV when $N_{\varphi}=16$
and -1.86eV when $N_{\varphi}=8$ while for $Si_{353}H_{196}$ it
is $-3.32eV$ for both $N_{\varphi}=16$ and $N_{\varphi}=8$. %
}. This suggests that the computational complexity (storage and computational
time) of the sTDH (or sTD-DFT) step scales linearly with system size. 

\begin{figure}[t]
\begin{centering}
\includegraphics[width=8cm]{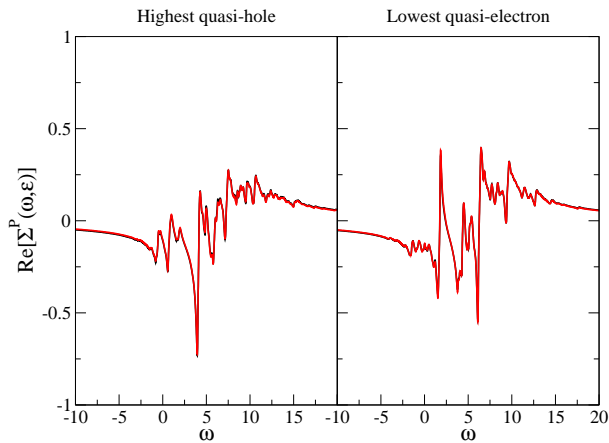}
\par\end{centering}

\caption{\label{fig:validation}Comparison of the stochastic (red) and deterministic
(black) estimates of the real part of the polarization self-energy
$\tilde{\Sigma}^{P}\left(\omega,\varepsilon\right)$ for the 14 electron
benchmark model corresponding to the highest quasi-hole and lowest
quasi-electron levels. Frequency scale in arbitrary units.}
\end{figure}

We validate our formalism by first applying it to a small model system
where a deterministic GW calculation is available as a benchmark~%
\footnote{The model system is a $\mbox{Si\ensuremath{H_{4}}}$ molecule represented
on a $4\times4\times4$ grid with a spacing of $1.25a_{0}$ and in
order to get a substantial HOMO-LUMO gap we filled the system with
14 electrons. Given the level of theory, i.e. $G_{0}W$ on top of
LDA and RPA screening the two methods make no further approximations.%
}. In Fig.~\ref{fig:validation} we show the estimates for the real
part of the polarization self-energy, obtained by both the deterministic
and the stochastic methods. The stochastic calculation employed a
large number of iterations ($I_{sGW}=10,000$), to achieve small statistical
errors. The agreement between the results of the two calculations
for all relevant frequencies as seen in Fig.~\ref{fig:validation}
is impressive for both the highest quasi-hole and lowest quasi-electron
levels, validating the stochastic formulation.

\begin{table}[H]
\caption{\label{tab:Parameters}The number of electrons ($N_{e}$), size of
grid ($N_{g}$), number of sDFT iterations ($I_{sDFT}$), number of
stochastic orbitals in sTD-DFT ($N_{\varphi}$), the value of $\beta_{GW}^{-1}$
($E_{h}$) in the sGW calculation, and the resulting QP energy gap
($E_{gap}^{QP}$) compared to $\mbox{G\ensuremath{W_{f}}}$ and $\mbox{\ensuremath{\Delta}SCF}$
calculations.}

\begin{centering}
\begin{tabular}{|c|c|c|c|c|c|c|c|c|}
\hline 
\multirow{2}{*}{System} & \multirow{2}{*}{$N_{e}$} & \multirow{2}{*}{$N_{g}$} & \multirow{2}{*}{$I_{sDFT}$} & \multirow{2}{*}{$N_{\varphi}$} & \multirow{2}{*}{$\beta_{GW}^{-1}$} & \multicolumn{3}{c|}{$E_{gap}^{QP}$ (eV)}\tabularnewline
\cline{7-9} 
 &  &  &  &  &  & sGW & $\mbox{G\ensuremath{W_{f}}}$ & $\mbox{\ensuremath{\Delta}SCF}$\tabularnewline
\hline 
\hline 
$\mbox{\textrm{{Si}}\ensuremath{{}_{35}\textrm{{H}}_{36}}}$ & $176$ & $60^{3}$ & $3000$ & $16$ & $0.020$ & $6.2$ & $7.0{}^{a}$ & $6.2^{a}$\tabularnewline
\hline 
$\mbox{\textrm{{Si}}\ensuremath{{}_{87}\textrm{{H}}_{76}}}$ & $424$ & $64^{3}$ & $1600$ & $16$ & $0.012$ & $4.8$ &  & \tabularnewline
\hline 
$\mbox{\textrm{{Si}}\ensuremath{{}_{147}\textrm{{H}}_{100}}}$ & $688$ & $70^{3}$ & $800$ & $16$ & $0.010$ & $4.1$ & $5.0^{a}$ & $4.1^{a}$\tabularnewline
\hline 
$\mbox{\textrm{{Si}}\ensuremath{{}_{353}\textrm{{H}}_{196}}}$ & $1608$ & $90^{3}$ & $400$ & $16$ & $0.008$ & $3.0$ &  & $2.9^{b}$\tabularnewline
\hline 
$\mbox{\textrm{{Si}}\ensuremath{{}_{705}\textrm{{H}}_{300}}}$ & $3120$ & $108^{3}$ & $200$ & $16$ & $0.007$ & $2.2$ &  & $2.4^{b}$\tabularnewline
\hline 
\end{tabular}
\par\end{centering}

(a) From Ref. \cite{Tiago2006}

(b) From Reg. \cite{Ogut1997}
\end{table}

Next we performed a set of sGW calculations for a series of hydrogen
passivated silicon NCs as detailed in Table~\ref{tab:Parameters}.
The sDFT method was used to generate the Kohn-Sham Hamiltonian within
the local density approximation (LDA). The calculations employed a
real-space grid of spacing $h=0.6a_{0}$, the Troullier-Martins norm-conserving
pseudopotentials~\cite{Troullier1991} and fast Fourier transforms
for implementing the kinetic and Hartree energies. The CPU time needed
to converge the sDFT to a statistical error in the total energy per
electron of about 10 meV was $\approx5000\,\mbox{hrs}$ for the entire
range of systems studied. 

\begin{figure}[t]
\begin{centering}
\includegraphics[width=8cm]{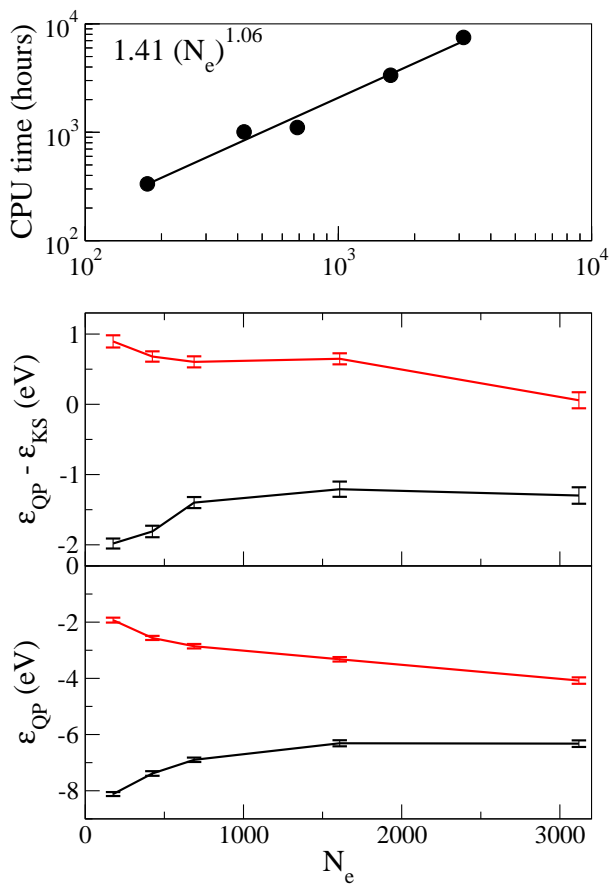}
\par\end{centering}

\caption{\label{fig:Si NC results} Lower panel: QP energies for the highest
quasi-hole (black) and lowest quasi-electron (red) levels. Middle
panel: QP energy difference from the KS energy for the highest quasi-hole
(black) and lowest quasi-electron (red) levels. Upper panel: CPU time
versus the number of electrons. The power law fit (solid line) yields
an exponent close to $1$.}
\end{figure}

In the lower panel of Fig.~\ref{fig:Si NC results} we plot the QP
energies of the highest quasi-hole and lowest quasi-electron levels
for the silicon NCs. We have used $I_{sGW}=1000$ stochastic iterations
and for each stochastic choice of $\phi$ and $\zeta$, we used $N_{\psi}=100$
stochastic $\psi$'s to generate the results. As can be seen, the
statistical error in the values of the QP energies is very small ($<0.1\,\mbox{eV}$)
and can be reduced by increasing $I_{sGW}$. The quasi-hole (quasi-electron)
energy increases (decreases) with system size due to quantum confinement
effect. The quasiparticle energies tend to plateau and approach the
bulk value as the size of the NC increases. The onset of the plateau
for electron seems to exceed the size of systems studied. This is
consistent with the fact that the effective mass of the electron is
smaller than that of the hole. The middle panel of Fig.~\ref{fig:Si NC results}
shows the QP energy difference from the KS values for the holes and
electrons. Larger deviations are observed for small NCs in the strong
confinement regime. The corrections for the holes are larger than
that for the electrons, which is rather surprising.

The upper panel of Fig.~\ref{fig:Si NC results} shows the scaling
of the entire sGW approach for the combined calculation of $\Sigma^{X}\left(\varepsilon\right)$,
$\Sigma^{XC}\left(\varepsilon\right)$, and $\Sigma^{P}\left(t;\varepsilon\right)$.
The scaling of the approach is nearly linear with the number of electrons,
breaking the quadratic theoretical limit. This near-linear scaling
behavior kicks in already for the smallest system studied and therefore
the stochastic method outperforms the ordinary $O\left(N^{4}\right)$
GW approach for all systems studied beyond $\mbox{Si}\mbox{\ensuremath{H_{4}}}$.
It is important to note that for almost the entire range of NC sizes
the sGW calculations were cheaper than the sDFT. 

We have also tested the sGW performance on PCBM (Phenyl-C61-butyric
acid methyl ester), a large non-symmetric system. We obtained $\varepsilon_{QP}=7.1\pm0.1$~eV
for the hole and $\varepsilon_{QP}=3.4\pm0.1$~eV for the electron
using $I=600$ iterations. These results can be compared to the experimental
results $E_{IP}=7.17$~eV and $E_{EA}=2.63$~eV \cite{larson2013electron,Akaike2008}.
The agreement for the electron affinity can be improved by replacing
the RPA screening with TDDFT screening \cite{Godby1994}, which gives
$\varepsilon_{QP}=2.5\pm0.1$~eV for the electron. The error per
iteration is thus similar to that of the symmetric silicon nanocrystalline
systems.

In conclusion, we have reformulated the GW approximation to MBPT for
QP energies as a stochastic process without directly referring to
KS eigenstates (or, equivalently, the single-particle density matrix).
sGW is a fully quantum paradigm shift and removes the main obstacle
for addressing large systems up to the mesoscopic limit. Indeed, the
application to silicon NCs of size far exceeding the current state-of-the-art
indicates that the complexity is near linear with system size, breaking
the theoretical limit. Some of the concepts presented here may be
applicable to other forms of MBPT, such as propagator \cite{linderberg2004propagators}
and Green's function theories \cite{cederbaum1977theoretical}. 

The sGW developed here has several appealing advantages:
\begin{itemize}
\item Representation: It is especially suitable for real-space-grid/plane-waves
pseudopotential representations for which the Hamiltonian operation
on a stochastic orbital scales linearly. These representations are
the natural for large-scale electronic structure computations. The
approach is also be useful for periodic systems with very large super-cells. 
\item CPU time scaling: The present method enables a GW calculation that
scales near-linearly in CPU time. Existing methods have been able
to reduce the complexity to cubic and it was implicitly assumed that
linear scaling is impossible due to the complexity of RPA. The present
method circumvents this by developing sTDH. The scaling of our approach
is insensitive to the sparsity of the density matrix and thus represents
a significant improvement over existing GW implementations. 
\item Storage scaling (matrix compression): The introduction of the stochastic
orbitals circumvents the need to store huge matrices of the Green
function and the polarization potential (or the inverse dielectric
matrix $\epsilon^{-1}$ etc.) thus achieving considerable savings
in memory. The scaling of storage is $O\left(N_{g}\right)$, which
makes sGW applicable to large system without recourse to various energy
cutoff approximations in the unoccupied space~\cite{Rieger1999,Samsonidze2011,Berger2012}.
\item Parallelization: The stochastic character of the sGW allows for straightforward
parallelization: self-energies are averaged over different stochastic
orbitals and each processor performs its own independent contribution
to this average.
\end{itemize}
These features make sGW the method of choice for studying QP excitations
in large complex materials not accessible by other approaches. 

R. B. and E. R. gratefully thank the Israel Science Foundation, Grants
No. 1020/10 and No. 611/11, respectively. R. B. and D. N. acknowledge
the support of the US-Israel Bi-National Science Foundation. D. N.,
Y. G., C. A. and C. K. are part of the Molecularly Engineered Energy
Materials (MEEM), an Energy Frontier Research Center funded by the
DOE, Office of Science, Office of Basic Energy Sciences under Award
No. DE-SC0001342.  Some of the simulations used resources of the Argonne
Leadership Computing Facility at Argonne National Laboratory, which
is supported by the Office of Science of the U.S. Department of Energy
under contract DE-AC02-06CH11357.

\bibliographystyle{apsrev4-1}
\bibliography{biblib}

\end{document}